\documentclass[runningheads]{llncs}
\usepackage{cite}
\usepackage{amsmath,amssymb,amsfonts}
\usepackage{graphicx}
\usepackage{textcomp}
\usepackage{soul} 

\usepackage{textcomp}
\usepackage{xcolor}
\usepackage{rotating}
\usepackage{array}
\usepackage{multirow}
\usepackage{subcaption}
\usepackage[colorlinks=true, linkcolor=blue, citecolor=blue, urlcolor=blue]{hyperref}
\usepackage{booktabs}
\usepackage{pdflscape}
\usepackage{float}
\usepackage[noend]{algpseudocode}
\usepackage{algorithm}
\usepackage{comment}
\begin{document}
\title{Engineering Attack Vectors and Detecting Anomalies in Additive Manufacturing\thanks{Supported by NSF - National Science Foundation}}
%
%
\author{Md Mahbub Hasan\inst{1}\orcidID{0009-0002-2226-6494} \and
Marcus Sternhagen\inst{1}\orcidID{0009-0001-5605-6631} \and
Krishna Chandra Roy\inst{1}\orcidID{0000-0001-9388-8042}}
\authorrunning{Hasan et al.}
%
\institute{New Mexico Institute of Mining and Technology\\
\email{\{mdmahbub.hasan,marcus.sternhagen\}@student.nmt.edu}, \email{krishna.roy@nmt.edu}}
\maketitle              
\begin{abstract}
Additive manufacturing (AM) is rapidly integrating into critical sectors such as aerospace, automotive, and healthcare. However, this cyber-physical convergence introduces new attack surfaces, especially at the interface between computer-aided design (CAD) and machine execution layers. In this work, we investigate targeted cyberattacks on two widely used fused deposition modeling (FDM) systems, Creality's flagship model K1 Max, and Ender 3. Our threat model is a multi-layered Man-in-the-Middle (MitM) intrusion, where the adversary intercepts and manipulates G-code files during upload from the user interface to the printer firmware. The MitM intrusion chain enables several stealthy sabotage scenarios. These attacks remain undetectable by conventional slicer software or runtime interfaces, resulting in structurally defective yet externally plausible printed parts. To counter these stealthy threats, we propose an unsupervised Intrusion Detection System (IDS) that analyzes structured machine logs generated during live printing. Our defense mechanism uses a frozen Transformer-based encoder (a BERT variant) to extract semantic representations of system behavior, followed by a contrastively trained projection head that learns anomaly-sensitive embeddings. Later, a clustering-based approach and a self-attention autoencoder are used for classification. Experimental results demonstrate that our approach effectively distinguishes between benign and compromised executions. 

\end{abstract}
\section{Introduction}
Additive Manufacturing (AM), commonly known as 3D printing, has revolutionized modern manufacturing by enabling the rapid prototyping and production of complex components with minimal material waste\cite{article2023, abdulhameed2019additive}. Its integration into safety-critical domains such as aerospace, healthcare, automotive, and defense has made the underlying infrastructure of AM systems a prime target for cyber-physical threats\cite{206156,7827651}. With the proliferation of network-connected 3D printers\cite{7958579}, adversaries can exploit attack vectors across the digital thread from CAD design, STL/G-code translation, network communication, to firmware execution, potentially sabotaging the mechanical integrity, dimensional accuracy, or intellectual property (IP)\cite{8742628} of printed parts~\cite{DBLP:journals/corr/abs-2111-12746,10.1145/2976749.2978300}. Recent studies have demonstrated the feasibility and consequences of attacks at various stages of the AM pipeline. These include manipulations of CAD/STL files~\cite{STURM2017154}, malicious firmware modifications~\cite{298969, inproceedings2017}, G-code-level sabotage~\cite{rosselsecurity}, and side-channel IP leakage through power, acoustic, and magnetic emissions\cite{7479068, 10.1145/3471621.3471850, 10.1145/2976749.2978300,8742628}. Meanwhile, attack detection frameworks have been proposed using process monitoring~\cite{10.1145/3264918, 10190501}, statistical modeling~\cite{DBLP:journals/corr/abs-2111-12746}, and analog emission analysis~\cite{7827651}. Despite these efforts, most prior approaches\cite{mahmood2025novel} rely on either predefined attacker models or static analysis or require access to a golden reference model(STL).
Hence, a significant gap persists in detecting real-time, stealthy attacks that operate directly at the G-code level, the machine-readable instruction. G-code is generated by slicer software from a CAD model and executed during fabrication. Once uploaded, the assumption of trust in the G-code pipeline creates a critical vulnerability. Unlike STL manipulations, G-code-level attacks can subtly alter toolpaths, extrusion volumes, or print timing without triggering visual inspection or violating basic geometry constraints. 

In this work, we explore stealthy G-code manipulation strategies that bypass traditional STL-based validation and compromise the final print without overt disruption. We identify three strategies that exploit realistic threat models in networked 3D printing setups, \textbf{Deferred Print Exploit}, \textbf{Access-Jammed G-code Swap}, and \textbf{Execution-Phase Tampering}.
Each of these approaches was implemented in a realistic threat model, such as Under-extrusion, Over-extrusion, Noisy G-code Injection, Dimensionality Change, and Internal Cavity Insertion. Assuming the adversary has access to the printer’s file system or control interface (e.g., compromised print servers, remote access tools, or insider threats). We conducted experiments on two widely used FDM platforms, Creality K1 Max and Creality Ender 3, with distinct system architectures (Klipper and OctoPrint, respectively), demonstrating the feasibility and cross-platform generalizability of these threats.

A central challenge lies in detecting such attacks without any ground truth STL or reference model, especially when deviations are subtle (e.g., minor under-extrusion or dimensional drift). To address this, we propose a log-based intrusion detection system (IDS) that operates in an unsupervised manner, trained solely on benign printer logs. Our system extracts telemetry and execution logs from the printer. 
We then utilize self-supervised contrastive learning with a frozen "MiniLM Transformer" encoder to generate discriminative latent embeddings. 
This approach preserves semantic structure without requiring annotated attack data. The embeddings are first visualized and segmented using PCA/UMAP and K-Means clustering to confirm separation between benign and anomalous patterns. For fine-grained anomaly detection, we further train a self-attention-based autoencoder on benign embeddings and measure reconstruction error during inference. Samples with high reconstruction loss are flagged as anomalies. Our key contributions are as follows:\\
\noindent\textbf{Realistic Attack Scenarios:} We construct multiple realistic adversarial scenarios based on a custom-
developed a backdoor interface that mimics legitimate printer control panels. These attacks include Under-extrusion, Over-extrusion, Noisy G-code Injection, Dimensionality Change, and Internal Cavity Insertion.

\noindent\textbf{Log-Centric IDS:} 
Our approach utilizes structured telemetry logs natively produced by the printer firmware. This allows
for non-invasive and real-time anomaly detection. 

\noindent \textbf{Representation Learning:} We employed a pretrained and frozen MiniLM language model to encode into semantically meaningful embeddings. A contrastive projection head is trained using only benign logs, enabling the model to learn a discriminative latent space that highlights deviations without requiring labeled attack data or golden STL/G-code references.

\noindent \textbf{Hybrid Evaluation:} We evaluate anomaly separation using both unsupervised clustering and reconstruction-based detection. A self-attention-based autoencoder is trained on benign embeddings to reconstruct normal patterns, and anomalies are flagged via high reconstruction loss.

Our approach complements prior literature by shifting focus from static design-level validation to dynamic runtime anomaly detection. 

\section{Background and Preliminaries}
\subsection{3D Printing Workflow}

\begin{figure}[h]
    \centering
    \begin{subfigure}[a]{0.45\textwidth}
        \centering
        \includegraphics[width=\textwidth]{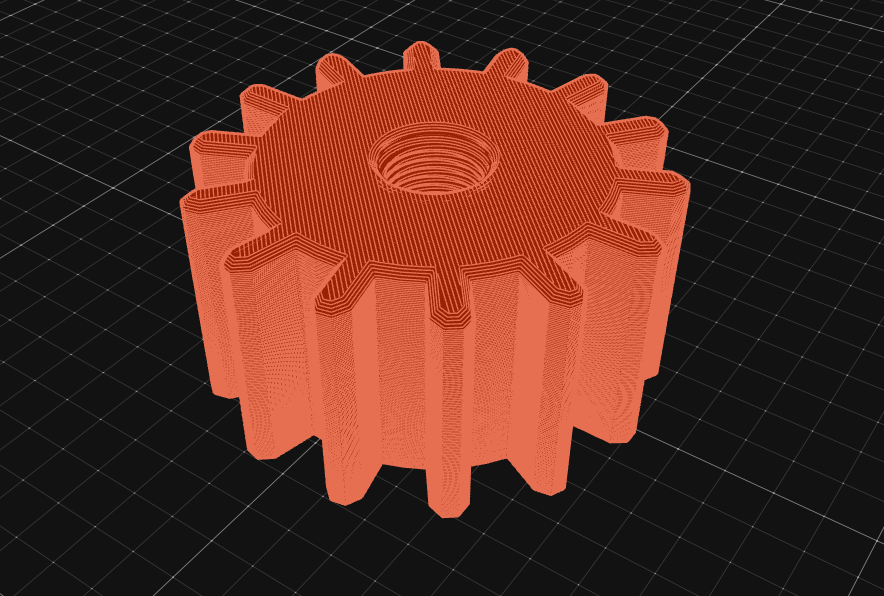}
        \caption{}
        \label{fig:Stereolithography}
    \end{subfigure}
    \hfill 
    \begin{subfigure}[a]{0.45\textwidth}
        \centering
        \includegraphics[width=\textwidth]{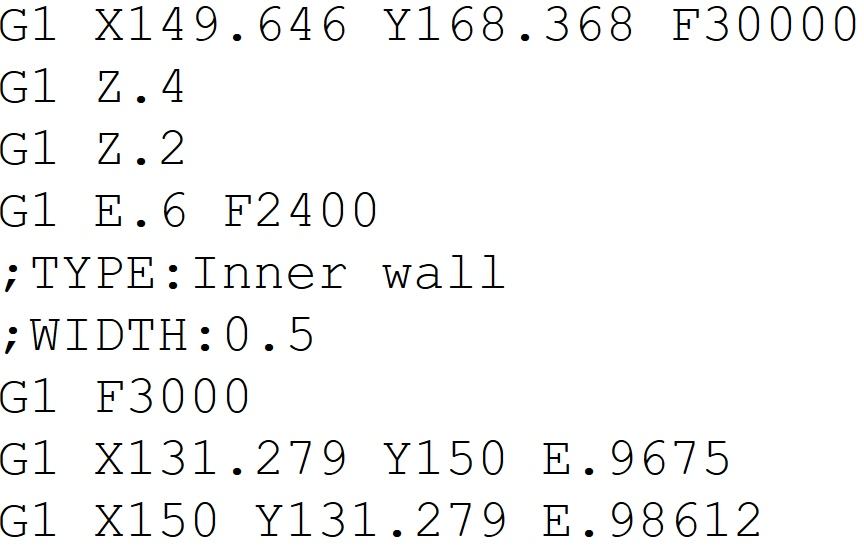}
        \caption{}
        \label{fig:gcode}
    \end{subfigure}
    \hfill 
    \caption{(a) Stereolithography Render, (b) G-code structure.}
    \label{fig:two_figures}
\end{figure}

3D printing is a layer-by-layer fabrication process that converts digital models into physical parts. Among various AM techniques, Fused Deposition Modeling (FDM)\cite{SOLOMON2021509} is the most prevalent due to its affordability, accessibility, and simplicity. In FDM, a thermoplastic filament\cite{mishra2023fdm} is melted and extruded through a heated nozzle, which moves across X, Y, and Z axes to deposit material onto a build platform. The process repeats layer by layer until the entire object is formed\cite{https://doi.org/10.5402/2012/208760}. A 3D printing workflow begins with a digital 3D model designed in computer-aided design (CAD) software. This model is exported in a mesh-based format, most commonly the STL (Stereolithography) file format\cite{PAUL201586}, which encodes the surface geometry of the object, Fig~\ref{fig:Stereolithography}. The STL file is then processed by a slicer, a software tool that slices the 3D model into discrete horizontal layers and converts them into machine-readable instructions. These instructions are encoded as G-code\cite{RAJAGURU2020628}, which defines precise commands for printer motion, extrusion, temperature control, and timing. The G-code file is typically uploaded to the printer and executed line-by-line during the manufacturing process, as demonstrated in Fig~\ref{fig:gcode}.

In our research, we investigated two widely used FDM printers, the Creality K1 Max and the Creality Ender 3, each configured with distinct system architectures and control interfaces.

\subsection{K1 Max \& Ender 3: Architecture}

The K1 Max runs a Klipper-based firmware and is connected to a custom Linux-based host via web interfaces such as Mainsail and Fluidd. Under the hood, Klipper offloads G-code parsing and motion planning to the host machine, which then communicates with the printer’s microcontroller via a serial interface. This architecture significantly improves performance and flexibility while exposing the communication channel between the host and MCU as a potential attack surface. To support real-time control and API access, Klipper is paired with Moonraker, a web service that provides a JSON-RPC interface for interacting with the printer state. Users upload G-code files through web frontends like Fluidd or Mainsail, which send the file to the Moonraker API, store it in a directory accessible to Klipper, and initiate printing through API calls. This separation between UI, API layer, and firmware creates multiple interception points for attackers who can gain access to the local network or file system.

In contrast, our Creality Ender 3 printer operates with a Raspberry Pi-based OctoPrint setup. In this configuration, the OctoPrint software runs on the Raspberry Pi and interfaces directly with the printer via USB. Users access a web dashboard hosted by OctoPrint to upload G-code files, monitor temperature and status, and initiate prints. OctoPrint acts as both the controller and relay, forwarding commands directly to the printer firmware. While simpler than the K1 Max architecture, OctoPrint still introduces a central point where G-code files can be intercepted or manipulated before execution.

\subsection{Attack Surface}
In our threat model, the attacker is positioned between the user and the firmware either by compromising the local network, gaining access to the proxy interface (e.g., Fluidd, Moonraker, or OctoPrint), or directly manipulating the file system. We implemented multiple stealthy MitM attacks that operate at the G-code layer, as demonstrated in Table~\ref{tab:relevant_gcode_attacks}. 

\begin{table}[ht]
\small
\centering
\caption{G-code Categories Used in Implemented Attack Scenarios}
\label{tab:relevant_gcode_attacks}
\begin{tabular}{|p{5cm}|p{5cm}|}
\hline
\textbf{Strategy} & \textbf{Relevant G-Codes} \\
\hline
\text{Deferred Print Exploit} & M25, M0, G4, G28, G92 \\
\text{Access-Jammed G-code Swap} & G92, M28, M23, M24, M25 \\
\text{Execution-Phase Tampering} & G4, M25, M24, G92, G1, G0 \\
\hline
\textbf{Attack Type} & \textbf{Relevant G-Codes} \\
\hline
\text{Under-Extrusion Attack} & G0, G1, M83, M221, M404 \\
\text{Over-Extrusion Attack} & G0, G1, M82, M221, M404 \\
\text{Noise Injection } & G0, G1, G92, M104, M109 \\
\text{Cavity Insertion} & G0, G1, G92, M28 \\
\text{Dimensional Manipulation} & G92, M206, G1, G92.X, M579 \\
\hline
\text{Intellectual Property Theft} & M154, M928, M111 \\
\hline
\end{tabular}
\end{table}

\section{Related Work}



Early works have demonstrated how 3D design files, such as STL, can be manipulated to introduce structural weaknesses. Logan et al.\cite{STURM2017154} showed how subtle changes to STL files could evade human inspection and lead to mechanical failure. Similarly, Sofia et al.\cite{206156} presented an end-to-end sabotage attack that compromised a drone propeller by manipulating design files, causing it to fail mid-flight. These studies highlight the potential of undetected design modifications to undermine the integrity of printed parts.
As an intermediate machine-readable format, G-code has become a focal point for adversaries. Caleb et al.\cite{DBLP:journals/corr/abs-2111-12746} explored subtle malicious edits to G-code without access to golden models. Their red-team/blue-team setup quantified the difficulty of detecting fine-grained G-code manipulations. "SOK" \cite{inproceedings2017} expanded on this by analyzing 278 potentially harmful G-codes and introducing new attacker models that can exploit even limited access. To defend against such threats, research has proposed reverse engineering G-code to validate printed geometries. Tsoutsos et al.~\cite{10.1145/3055186.3055198} developed a toolpath reconstruction method to re-generate approximate 3D models from G-code and assess structural soundness via Finite Element Analysis (FEA).

Firmware-level compromises have been shown to offer persistent and stealthy attack vectors. Mahmood et al.\cite{mahmood2025novel} demonstrated that malicious firmware could intercept or replace print jobs and manipulate extrusion settings. Muhammad et al.\cite{298969} systematized firmware attacks across multiple stages of 3D printing and introduced an "Attack Feasibility Index" (AFI) to rank potential threats. Beyond firmware, side-channel attacks leveraging acoustic, electromagnetic, or power emissions have also been explored. Song et al.\cite{10.1145/2976749.2978300} showed how smartphone sensors could reconstruct G-code by analyzing acoustic and magnetic data. "Encryption is Futile" article\cite{10.1145/3471621.3471850} further demonstrated that power side-channel leakage could yield near-perfect model reconstructions even under encrypted communications.
The risk of IP leakage has driven efforts to secure AM pipelines. Yampolskiy et al.\cite{10.1145/2689702.2689709} proposed secure outsourcing models for IP-sensitive manufacturing. In contrast, Chhetri et al.\cite{8742628} highlighted how modifying compilers could amplify side-channel leakage to leak G-code content. On the network side, McCormack et al. \cite{mccormackc3po,9283836} introduced C3PO, a tool for assessing security posture and attack paths in networked 3D printing deployments.
Real-time process monitoring is another line of defense. Gao et al.\cite{10.1145/3264918} developed a sensor-based framework for monitoring physical parameters such as speed, infill, and fan speed to detect abnormal printing behavior. "KCAD" paper\cite{7827651} introduced a kinetic cyber-attack detection system using statistical mappings between analog emissions and G-code patterns. Recent work by Mahmood et al.\cite{mahmood2025novel} and Meleshko et al.\cite{10974798} focuses on modeling cyber-attacks in laboratory setups, offering simulation-based evaluations of malicious modifications and their detection.

Unlike prior works that focus on a single attack vector, such as G-code tampering, side-channel exfiltration, or firmware backdoors, our approach provides a comprehensive threat model that encompasses multiple intrusion points within the 3D printing workflow. We design a realistic attacker model that utilizes a rogue web interface to intercept user activity, manipulate uploaded G-code files pre- and mid-print, and hijack printer controls. 
Additionally, we propose an unsupervised anomaly detection system that uses structured printer host logs and transformer-based log embedding with contrastive learning to differentiate benign vs. manipulated prints.


\section{Attack Strategy}

\subsection{Assumptions}

Our threat model assumes that the attacker has obtained root-level access to the 3D printer's host operating system by compromising authentication credentials. 
The assumption is realistic and supported by both prior observations of poor authentication practices in embedded printing systems and our experimental validation.

In a controlled lab environment, we validated the feasibility of credential compromise using both brute-force and dictionary-based password attacks. These experiments were conducted on Creality K1 Max and Ender 3 setups, where SSH rate limiting was disabled and weak passwords were used to simulate real-world misconfigurations. The brute-force attack employed parallel enumeration of common character patterns and successfully broke short passwords frequently used in embedded systems. The dictionary-based approach further demonstrated effectiveness by leveraging a curated list of domain-specific keywords, including printer models (e.g., "Ender", "Creality"), materials (e.g., "PLA", "ABS"), and software terms (e.g., "OctoPrint", "Slicer", "Klipper"). These wordlists simulate real-world attacker strategies based on contextual familiarity with additive manufacturing environments. The success of these attacks aligns with findings from broader security surveys indicating that many 3D printers are deployed with factory-default credentials, passwords derived from printer names, or otherwise weak authentication schemes. Additionally, the embedded and unattended nature of many 3D printing controllers leads to inconsistent patching and poor security hygiene. Firmware-level authentication mechanisms are either absent or insufficient, and the presence of network-facing control interfaces (e.g., Moonraker, OctoPrint, or Fluidd) introduces an expanded attack surface. These services are often accessible over the LAN or, in some cases, exposed to the public internet for remote printing convenience, further enabling remote adversaries.

Establishing credential compromise as a feasible and reproducible step allows us to explore downstream attack scenarios that arise once root access is obtained. 

\subsection{Rogue Interface and Attacks}
One of the core intrusion techniques developed in this work involves the creation of a rogue control interface that mimics the legitimate web interface used for controlling 3D printers running printer firmware. This malicious replica is hosted on the attacker’s own machine or a compromised device within the same network as the target printer. Once root access is obtained through credential compromise, the attacker installs this mimic interface to act as a transparent proxy between the legitimate user and the underlying printer system.

It communicates directly with the print server API, accessing the printer’s file system and configuration directories. As a result, the attacker gains real-time visibility into all user actions and machine operations, including G-code uploads, printing progress, system temperatures, logging activity, and configuration changes. 
By deploying this replica interface as a stealth access layer, the attacker effectively bypasses traditional network-layer monitoring tools. This enables a suite of downstream G-code manipulation attacks with minimal detection risk, laying the foundation for the stealthy sabotage scenarios described in the following paragraphs.

\begin{figure}[h]
    \centering
    \includegraphics[width=.8\textwidth]{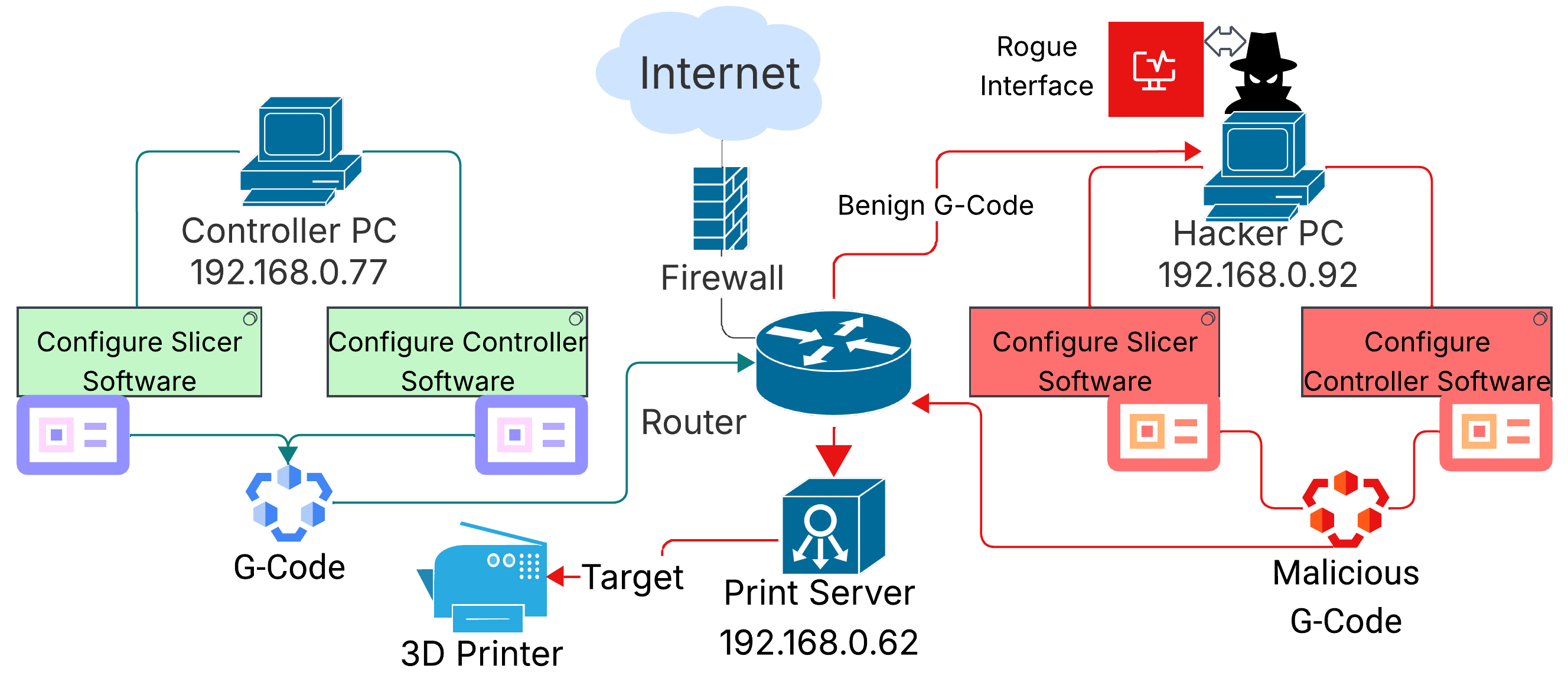}
    \caption{Attack Strategy Pipeline}
    \label{fig:mitm}
\end{figure}

\noindent\textbf{Threat Model:}
Our threat model assumes an adversary with a primary attack point into the AM pipeline,  which is system-level backdoor access to the printer’s file system and control interface. 
Fig~\ref{fig:mitm} illustrates the complete intrusion architecture. 

\paragraph{Intrusion Point: MitM with G-code Manipulation.}
The adversary utilizes unauthorized access to the printer's internal file system and G-code control flow via SSH and a spoofed web interface. We implement the following techniques:\\
\textbf{Deferred Print Exploit:}
The adversary monitors the printer’s G-code upload directory and waits for a legitimate user to upload a file via the standard web interface (e.g., Fluidd, Mainsail, or OctoPrint). Once a new .gcode file is detected, the attacker allows a short delay window to pass, ensuring the file is not being actively used by the system. After this short period, the attacker downloads the uploaded file, modifies it locally by injecting stealthy perturbations, and re-uploads the modified file to the original path. This form of post-upload interception is complex for users to detect, as most 3D printer UIs do not validate the integrity of uploaded files or retain cryptographic hashes\cite{DBLP:journals/corr/abs-2103-06400}. The entire operation takes place between the time the file is uploaded and before the user initiates printing. 
This strategy targets the window of vulnerability between upload and execution. 

\noindent\textbf{Access-Jammed G-code Swap:}
This attack refines the delayed replacement strategy by locking the uploaded G-code file immediately upon detection. As soon as a new .gcode file appears, the attacker temporarily makes it inaccessible to the user for 5 to 10 seconds. This ensures the file will not be executed prematurely while it is being modified. After locking, the attacker downloads the file, injects targeted modifications, and uploads the altered version. The file is then restored to its original name, completing the substitution. This method ensures fast replacement of the original file, 
and preventing user access during the manipulation process. It also increases the likelihood of successful sabotage even in environments with fast user-triggered printing.

\noindent\textbf{Execution-Phase Tampering:}
The adversary targets an active print job that is already underway. By querying the printer state via the print server API, the attacker determines the filename currently being executed and the print state. After a configurable delay, typically 2 to 5 minutes, the attacker intercepts the ongoing print using a direct command to the backend control interface. 
Then, the attacker replaces the currently printing G-code file with a modified version in the remaining layers. 
This approach allows attackers to target only the latter portion of the print, making anomalies less noticeable during early inspection and potentially bypassing surface-level quality control. It also avoids suspicion from the user, who may attribute final-layer defects to slicing issues or mechanical artifacts.

\noindent\textbf{Attacks by Intrusion Point:} The adversary’s access to G-code control enables a broad spectrum of malicious modifications, ranging from subtle sabotage to outright intellectual property theft. 

\noindent\textit{Intellectual Property (IP) Theft.}
With access to the printer’s file system, the adversary can exfiltrate proprietary design assets such as uploaded G-code files or STL derivatives. These files can be intercepted before or after any modifications and covertly transmitted to an attacker-controlled server. Once obtained, the files can be reverse-engineered to extract valuable manufacturing information such as pathing logic, infill strategies, or fine-tuned process parameters. Furthermore, the stolen files may be replicated on remote 3D printers to produce counterfeit copies of the original design. 
Such breaches are especially damaging in high-stakes sectors like aerospace, biomedical devices, or defense manufacturing, where the leakage of proprietary IP can result in significant strategic and economic consequences.

\noindent\textit{Under-extrusion:} By reducing \texttt{E} values, the nozzle delivers insufficient material, leading to weak interlayer bonding, poor layer adhesion, and ultimately, reduced tensile strength of the part. Such defects may be internal and invisible to post-print visual inspection.

\noindent\textit{Over-extrusion:} Conversely, inflating \texttt{E} values causes excessive filament deposition, which results in dimensional inaccuracy, surface roughness, or printer head drag. This may damage surrounding layers or cause the actuator to overheat.

\noindent\textit{Noisy G-code Injection:}
Another stealthy sabotage approach involves injecting noisy motion commands, such as high-frequency or random \texttt{G0} (non-extruding move) and \texttt{G1} (extruding move) instructions. These can increase wear and tear on stepper motors by inducing high-speed oscillations, surface-level distortions by introducing non-uniform paths, or jagged trajectories. It can also cause micro-delays that desynchronize thermal control, leading to inconsistent material melting. 

\noindent\textit{Dimensionality Change:}
Subtle manipulation of coordinate scaling, such as multiplying the \texttt{X}, \texttt{Y}, or \texttt{Z} values in selected G-code segments, can introduce imperceptible but harmful geometric distortions. Scaling along the\texttt{X}/\texttt{Y} axes affects footprint and fitment tolerances. Scaling the \texttt{Z} axis affects layer height, which in turn impacts stacking behavior and vertical strength. Localized changes (e.g., only in the middle third of the print) evade dimensional validation tools that measure only the bounding box. Such attacks compromise precision-critical components (e.g., interlocking parts or aerodynamics-sensitive designs) while maintaining overall visual similarity.

\noindent\textit{Internal Cavity Insertion:}
The adversary may target the structural core of the object by introducing voids or cavities within specific layer ranges. Cavities are created by removing or skipping \texttt{E} extrusion values across predefined \texttt{Z}-height intervals. These modifications are made mid-print to avoid detection in pre-print preview or slicing validation. While the external geometry remains intact, the absence of infill or perimeters in these regions significantly degrades load-bearing capacity, causing silent failures under stress. This attack is hazardous in safety-critical parts, where internal defects are not easily visible but have a direct impact on reliability.

\begin{figure}[h]
    \centering
    \begin{subfigure}[a]{0.48\textwidth}
        \centering
        \includegraphics[width=.48\textwidth]{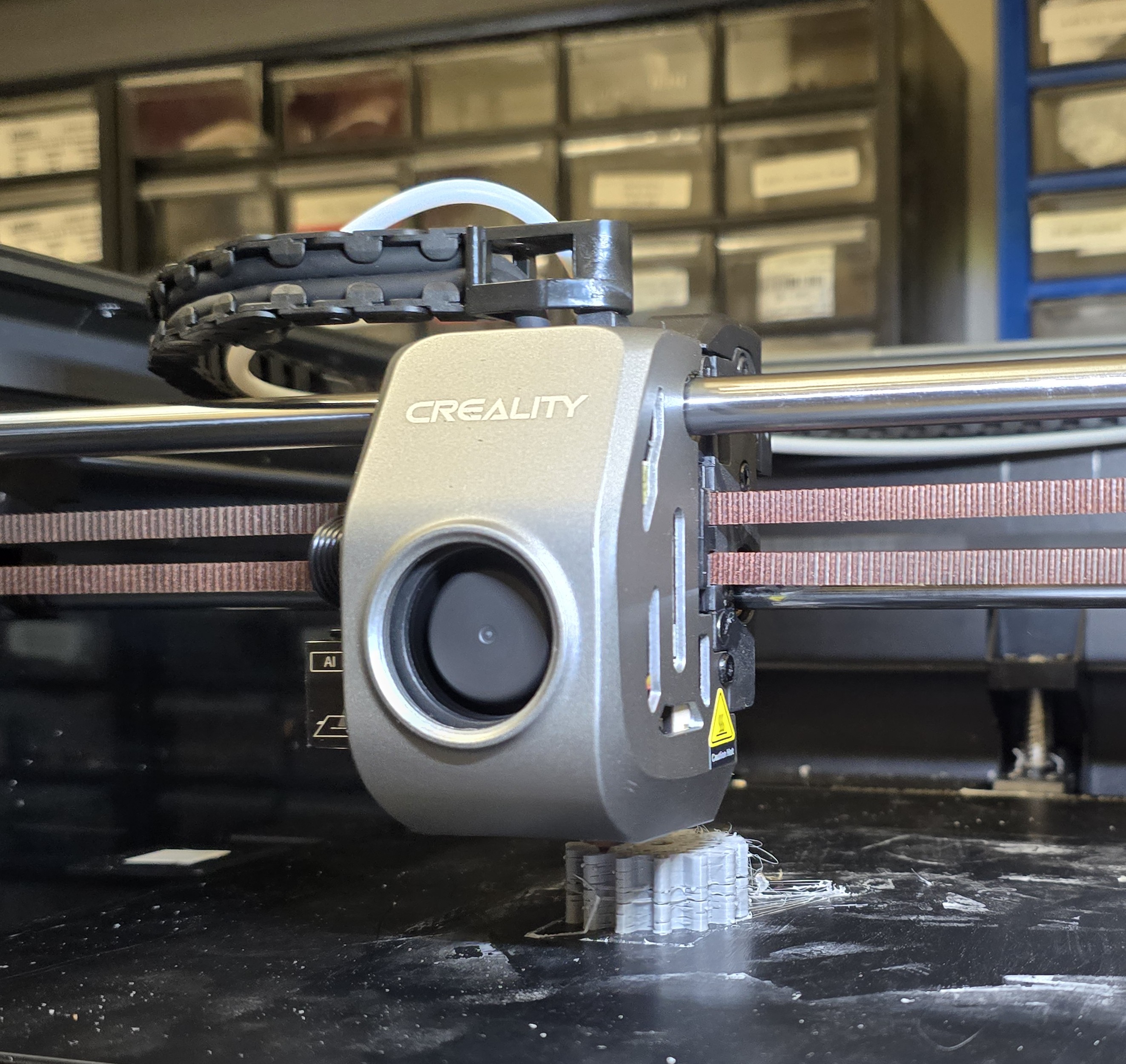}
        \caption{Noise Injected Object: K1-Max.}
        \label{fig:k1_print}
    \end{subfigure}
    \hfill 
    \begin{subfigure}[a]{0.48\textwidth}
        \centering
        \includegraphics[width=.48\textwidth]{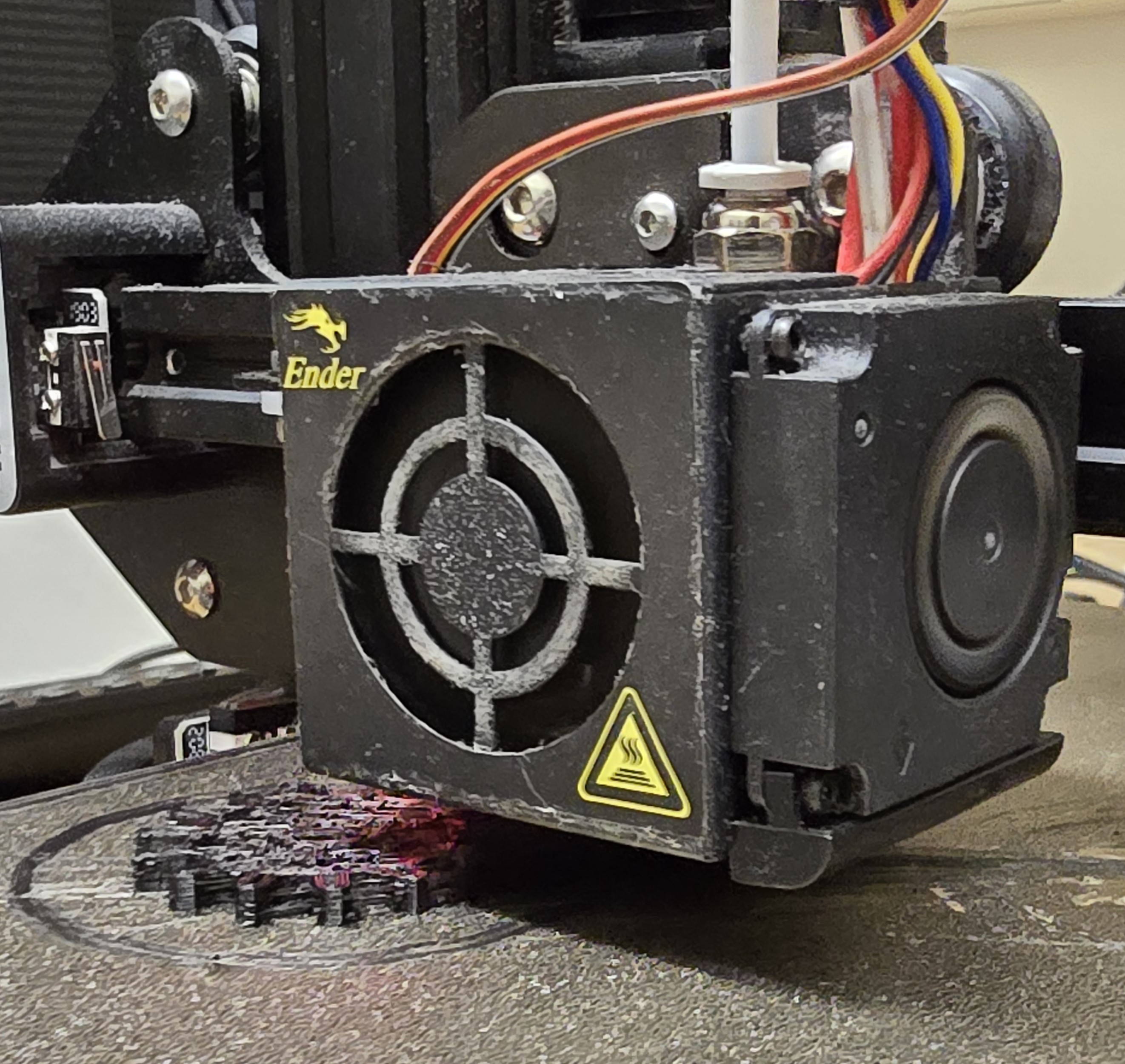}
        \caption{Cavity Inserted Object: Ender-3.}
        \label{fig:ender3_print}
    \end{subfigure}
    \hfill 
    \caption{K1-Max \& Ender-3 Printing Object During Attack.}
    \label{fig:K1Ender_print}
\end{figure}

\section{Defense Methodology}

\subsection{Data Collection}

We utilize printer logs as a high-fidelity data source to capture the operational state and runtime behavior of the 3D printer. 
The logs were collected under two primary conditions: (1) benign operation involving benign slicing and printing workflows, and (2) malicious scenarios. 

\subsection{Feature Selection}
To ensure that our model captures meaningful semantic patterns while minimizing redundancy, we employed a hybrid approach combining domain-driven selection and statistical filtering. Initially, we identified a broad set of features shown in the Table~\ref{tab:klipper_features} from logs based on their relevance to printing state, thermal stability, communication health, and system workload. To refine this selection, we applied a two-step statistical filtering pipeline:

\begin{table}[ht]
\small
\centering
\caption{Features from Logs}
\begin{tabular}{|p{2.5cm}|p{4cm}|p{5.2cm}|}
\hline
\textbf{Category} & \textbf{Feature(Raw logs)} & \textbf{Reason} \\
\hline
Print Progress & \texttt{print\_time}, \texttt{sd\_pos}, \texttt{buffer\_time}, \texttt{print\_stall}, \texttt{gcodein} & Captures how much of the print job has been completed and helps detect abnormal progression or interruptions. \\
\hline
Thermal Data & \texttt{mcu\_temp}, \texttt{chamber\_temp}, \texttt{heater\_bed.temp}, \texttt{extruder.temp}, \texttt{heater\_bed.pwm}, \texttt{extruder.pwm} & Reflects thermal stability; useful for identifying physical anomalies such as under-extrusion or overheating. \\
\hline
System Resources & \texttt{sysload}, \texttt{cputime}, \texttt{memavail} & Indicates CPU and memory stress. 
\\
\hline
MCU Stats & \texttt{bytes\_write}, \texttt{bytes\_read}, \texttt{bytes\_retransmit}, \texttt{bytes\_invalid}, \texttt{send\_seq}, \texttt{receive\_seq}, \texttt{retransmit\_seq} & Represents communication reliability between host and MCUs; abnormal retransmissions or packet loss may indicate attack. \\
\hline
MCU Timings & \texttt{mcu\_task\_avg}, \texttt{mcu\_task\_stddev}, \texttt{mcu\_awake}, \texttt{srtt}, \texttt{rttvar}, \texttt{rto} & Timing delays and task scheduling variance can highlight abnormal processing behavior or congestion. \\
\hline
Nozzle/Leveling MCU & \texttt{nozzle\_mcu.*}, \texttt{leveling\_mcu.*} & Comparing timing and communication across different MCUs can detect localized issues or attack traces. \\
\hline
Raspberry Pi Stats & \texttt{rpi.mcu\_task\_avg}, \texttt{rpi.bytes\_retransmit}, etc. & Shows host-controller interaction health; anomalies here can indicate local overload or network disruption. \\
\hline
\end{tabular}
\label{tab:klipper_features}
\end{table}

\noindent\textbf{Variance Thresholding:} Features with near-zero variance (threshold $<0.01$) across the benign dataset were discarded, as they offer limited discriminative power.\\
\noindent\textbf{Correlation Pruning:} From the remaining features, we computed a pairwise Pearson correlation matrix and eliminated features exhibiting high linear correlation ($>0.95$) to reduce redundancy.

\subsection{Model Architecture}

\begin{figure}[h]
    \centering
    \includegraphics[width=\textwidth]{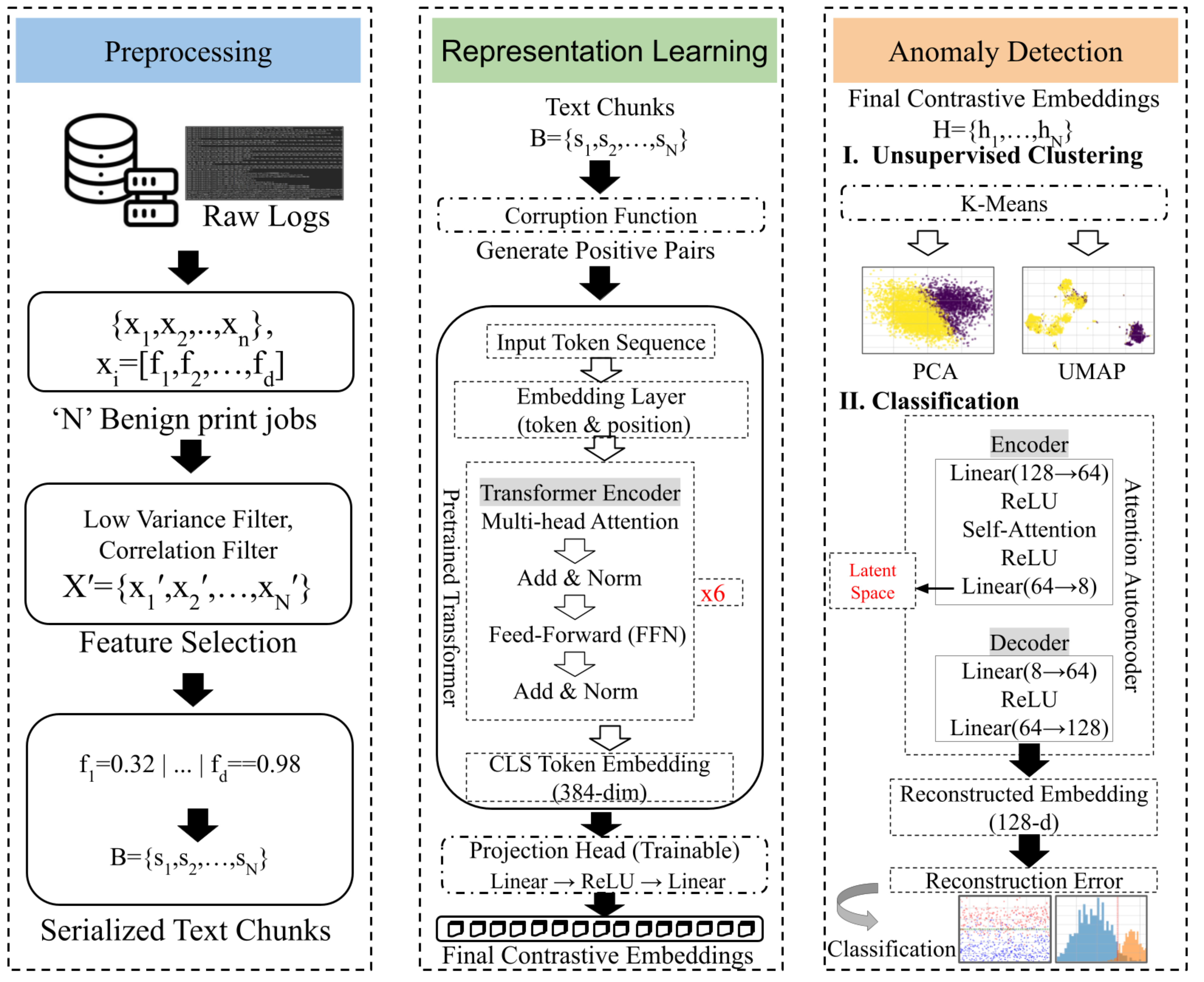}
    \caption{Model Architecture: Preprocessing, Representation Learning, and Anomaly Detection }
    \label{fig:model_architecture}
\end{figure}

We propose an unsupervised anomaly detection framework. 
The complete pipeline is demonstrated in the Figure~\ref{fig:model_architecture}. The architecture consists of two core stages: (i) contrastive representation learning using a frozen Transformer encoder, and (ii) anomaly detection using a self-attention-enhanced autoencoder.

The structured logs be represented as a matrix \(\mathcal{L} = \{ \mathbf{x}_1, \mathbf{x}_2, \ldots, \mathbf{x}_N \} \), where each row \(\mathbf{x}_i \in \mathbb{R}^d\) denotes a log with \(d\) numeric features. After applying low-variance and high-correlation filters, the reduced feature set is denoted as \(\mathcal{L'} = \{ \mathbf{x}'_1, \ldots, \mathbf{x}'_N \}\), where \(\mathbf{x}'_i \in \mathbb{R}^{d'}\), \(d' < d\). Each filtered row \(\mathbf{x}'_i\) is serialized into a sentence-like structure. This transformation preserves context and enables token-based encoding: $s_i = \text{"key}_1=v_1 \, | \, \text{key}_2=v_2 \, | \, \dots \, | \, \text{key}_d=v_{d'}"$

\noindent\textbf{Contrastive Representation Learning:} We employ the \texttt{all-MiniLM-L6-v2} transformer model as a frozen encoder \(f_\theta: \mathbb{S} \rightarrow \mathbb{R}^{384}\), where \(\theta\) represents the fixed pretrained parameters. For a given sentence \(s_i\), the frozen encoder generates a 384-dimensional semantic embedding: $\mathbf{h}_i = f_\theta(s_i),\ \mathbf{h}_i \in \mathbb{R}^{384}$


To specialize the embedding for anomaly detection, we introduce a trainable projection head \(g_\phi: \mathbb{R}^{384} \rightarrow \mathbb{R}^{128}\), parameterized by \(\phi\). This projection head is implemented as a double layer MLP: $g_\phi(\mathbf{h}) = \mathbf{W}_2 \cdot \text{ReLU}(\mathbf{W}_1 \cdot \mathbf{h} + \mathbf{b}_1) + \mathbf{b}_2$ ,


\noindent{with weight matrices \(\mathbf{W}_1 \in \mathbb{R}^{256 \times 384}\), \(\mathbf{W}_2 \in \mathbb{R}^{128 \times 256}\), and bias terms \(\mathbf{b}_1 \in \mathbb{R}^{256}\), \(\mathbf{b}_2 \in \mathbb{R}^{128}\).}
The complete model forward pass is defined as: $\mathbf{z}_i = g_\phi(f_\theta(s_i))$ where $\mathbf{z}_i \in \mathbb{R}^{128}.$
To train the model in an unsupervised manner, we adopt contrastive learning. For each benign line \(s_i\), we generate a synthetically perturbed version \(s_i'\) by applying domain aware corruption (e.g., replacing ``extruder'' with ``toolhead'', or modifying temperature labels). The goal is to minimize the contrastive loss between benign and corrupted embeddings: $\mathcal{L}_{\text{contrastive}} = 1 - \cos\left( \mathbf{z}_i, \mathbf{z}_i' \right) = 1 - \frac{\mathbf{z}_i \cdot \mathbf{z}_i'}{\|\mathbf{z}_i\| \cdot \|\mathbf{z}_i'\|}.$
Only the projection head parameters \(g_{\phi}\) are updated during training, while the base encoder \(f_\theta\) remains frozen. 

\noindent\textbf{Inference and Anomaly Detection:}
At inference time, unseen printer logs are serialized and passed through the frozen encoder \(f_\theta\) and the trained projection head \(g_\phi\) to generate compact embeddings: $\mathbf{z}_i = g_\phi(f_\theta(s_i)).$
These embeddings serve as inputs to two complementary anomaly detection strategies:

\paragraph{I. Clustering-based Detection:}
We apply K-Means clustering on benign embeddings to learn centroid representations of normal behavior. For a test embedding \(\mathbf{z}_i\), the anomaly score is computed as the minimum Euclidean distance to the nearest cluster centroid: $\mathcal{A}_i^{(\text{cluster})} = \min_{c \in C} \| \mathbf{z}_i - c \|_2.$
Embeddings with high cluster distance are considered anomalous. Dimensionality reduction techniques such as PCA and UMAP are also applied for visual validation of clustering boundaries.

\paragraph{II. Reconstruction-based Detection:}
To complement the clustering results, we utilize a self-attention-enhanced autoencoder to reconstruct embeddings generated from benign printer behavior. A high reconstruction error is indicative of a potential anomaly.
\(\mathbf{z}_i \in \mathbb{R}^{128}\) denote the contrastive embedding of a serialized log sentence. The autoencoder architecture \(\mathcal{A}_\psi\) consists of an encoder \(E\) and a decoder \(D\) such that: $\hat{\mathbf{z}}_i = D(E(\mathbf{z}_i)), \quad \text{for } i = 1, \ldots, N.$ \\
\textbf{Encoder Structure:} The encoder projects the input into a compressed latent space of dimension 8. This is achieved via:

\begin{align}
\mathbf{h}_i^{(1)} &= \text{ReLU}(W_1 \cdot \mathbf{z}_i + \mathbf{b}_1), \quad W_1 \in \mathbb{R}^{64 \times 128} \\
\mathbf{h}_i^{(attn)} &= \text{SelfAttention}(\mathbf{h}_i^{(1)}) \in \mathbb{R}^{64} \\
\mathbf{z}_i^{(enc)} &= W_2 \cdot \text{ReLU}(\mathbf{h}_i^{(attn)}) + \mathbf{b}_2, \quad W_2 \in \mathbb{R}^{8 \times 64}
\end{align}

\noindent\textbf{Self-Attention Mechanism:} We implement a single-head self-attention module to capture latent interactions among features within each embedding vector. Given an input \(\mathbf{h} \in \mathbb{R}^{64}\), the query, key, and value vectors are computed as\cite{NIPS2017_3f5ee243}: \( Q = W_Q \mathbf{h}, \; K = W_K \mathbf{h}, \; V = W_V \mathbf{h} \)
, where \(W_Q, W_K, W_V \in \mathbb{R}^{64 \times 64}\) are learnable weight matrices. The attention score is computed as: \( \alpha = \frac{Q \cdot K^T}{\sqrt{d_k}} \in \mathbb{R}, \; d_k = 64 \), and \( \text{AttentionOutput} = \text{softmax}(\alpha) \cdot V \)
. This operation allows the encoder to dynamically emphasize more informative feature components when forming the latent representation.

\noindent\textbf{Decoder Structure:} The decoder reconstructs the original 128-dimensional embedding from the latent code:
\begin{align}
\mathbf{h}_i^{(3)} &= \text{ReLU}(W_3 \cdot \mathbf{z}_i^{(enc)} + \mathbf{b}_3), \quad W_3 \in \mathbb{R}^{64 \times 8} \\
\hat{\mathbf{z}}_i &= W_4 \cdot \mathbf{h}_i^{(3)} + \mathbf{b}_4, \quad W_4 \in \mathbb{R}^{128 \times 64}
\end{align}

\paragraph{Anomaly Scoring:}
Reconstruction error is used to assign an anomaly score: \( \mathcal{E}_i = \frac{1}{128} \left\| \hat{\mathbf{z}}_i - \mathbf{z}_i \right\|_2^2 \)
A threshold \(\tau\) is determined from the 95th percentile of benign reconstruction errors. A test log is classified as anomalous if \(\mathcal{E}_i > \tau\).

\section{Experimental Analysis}

\subsection{Experimental Setup:}
The primary hardware platform used is the Creality K1 Max 3D printer, which operates on the Klipper firmware, interfaced through Moonraker, and accessed via the Fluidd and Mainsail web frontends. 
The attack automation is handled using Python 3.11, using 
SSH-based file access, requests for interacting with the Moonraker API, and an observer for real-time file system monitoring and G-code injection. In addition, Ender 3 is used with a Raspberry Pi 4 running OctoPrint to validate the portability of our attacks across different firmware and control stacks. Both printers were connected to the same local area network (LAN) as the attacker's host for seamless monitoring and command execution. All experiments were performed on a high-performance host machine equipped with an Intel Core i9-14900KF CPU clocked at 3.20 GHz and 32 GB of DDR5 RAM. 
We used an NVIDIA RTX 4090 GPU featuring 24 GB of GDDR6X memory, a 384-bit memory bus, and 512 Tensor cores optimized for AI workloads.

\subsection{Dataset Description}

Table~\ref{tab:sample_distribution} illustrates the size of each sample. All attack categories are deliberately excluded from training to simulate a realistic zero-day threat detection scenario. Class proportions are chosen to ensure diversity and avoid bias toward any specific attack.

\begin{table}[ht]
\small
\centering
\caption{Distribution for Training and Validation}
\begin{tabular}{lccc}
\toprule
\textbf{Class Type} & \textbf{Attack Category} & \textbf{\ Samples} & \textbf{Usage} \\
\midrule
Benign              & --                      & 98,720              & Training \\
Benign              & --                      & 57,373              & Validation \\
\midrule
\multirow{6}{*}{Attack} 
                   & Under-extrusion         & 10120                & Validation \\
                   & Over-extrusion          & 9950                & Validation \\
                   & Noise Injection         & 15456                & Validation \\
                   & Dimensional Change      & 9324                & Validation \\
                   & Cavity Insertion        & 10521                & Validation \\
\bottomrule
\end{tabular}
\label{tab:sample_distribution}
\end{table}

\subsection{Training Details}
A total of 98,720 benign samples were extracted from logs and structured into fixed-length time windows. The frozen MiniLM transformer model produces 384-dimensional dense embeddings for each log sequence. To adapt this dimension, we append a learnable multi-layer perceptron (MLP) projection head that maps the representations from 384 to 128 dimensions.



We use the Adam optimizer with a learning rate of $1 \times 10^{-4}$ and a batch size of 64. The model is trained for 25 epochs using only benign data, with no attack labels. 

\section{Qualitative Evaluation}

\setlength{\fboxsep}{.5pt}
\begin{figure}[h]
    \centering
    \subfloat[Benign.]{\fbox{\includegraphics[width=0.22\textwidth]{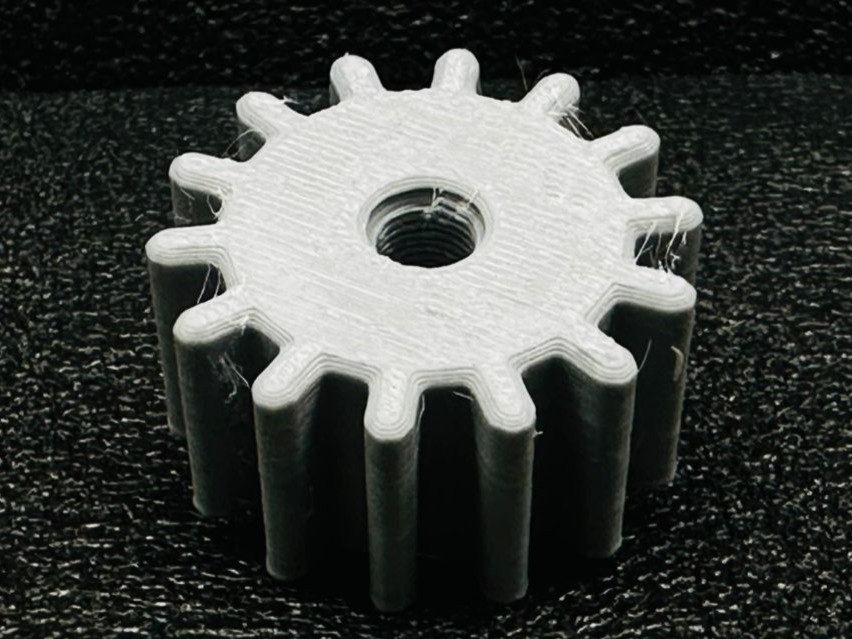}}} 
    \subfloat[Cavity Ins.]{\fbox{\includegraphics[width=0.22\textwidth]{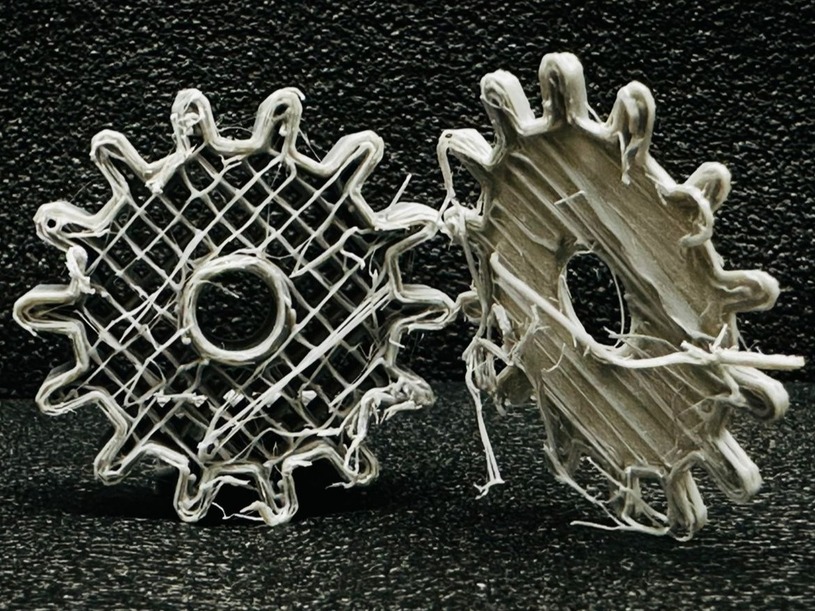}}} 
    \subfloat[Noise Inj.]{\fbox{\includegraphics[width=0.22\textwidth]{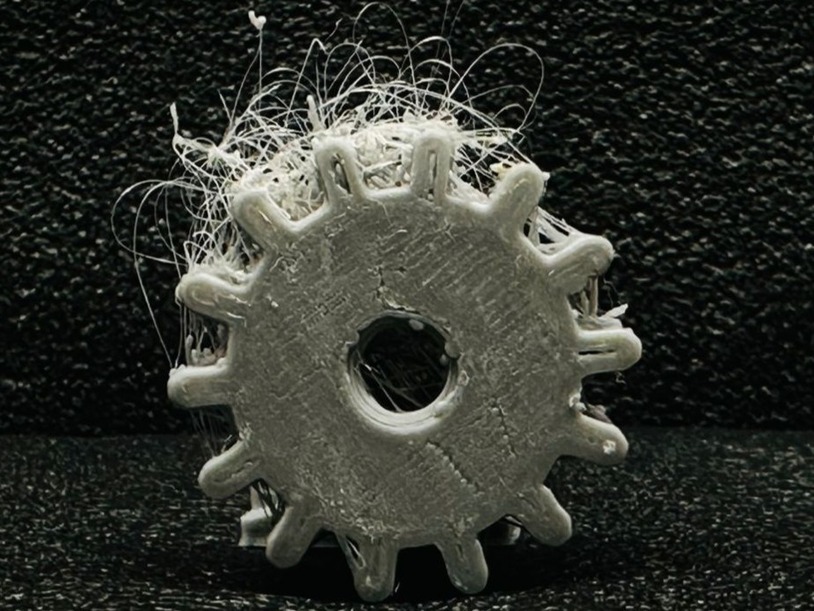}}} 
    \subfloat[Dimension Inj.]{\fbox{\includegraphics[width=0.22\textwidth]{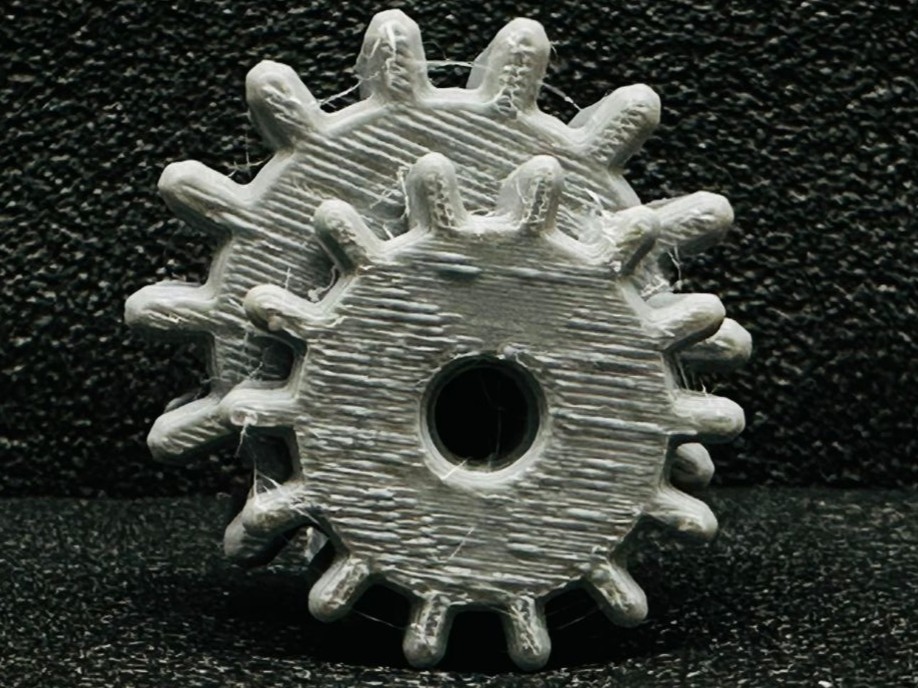}}}\\
    
    \subfloat[Benign Extrusion]{\fbox{\includegraphics[width=0.28\textwidth]{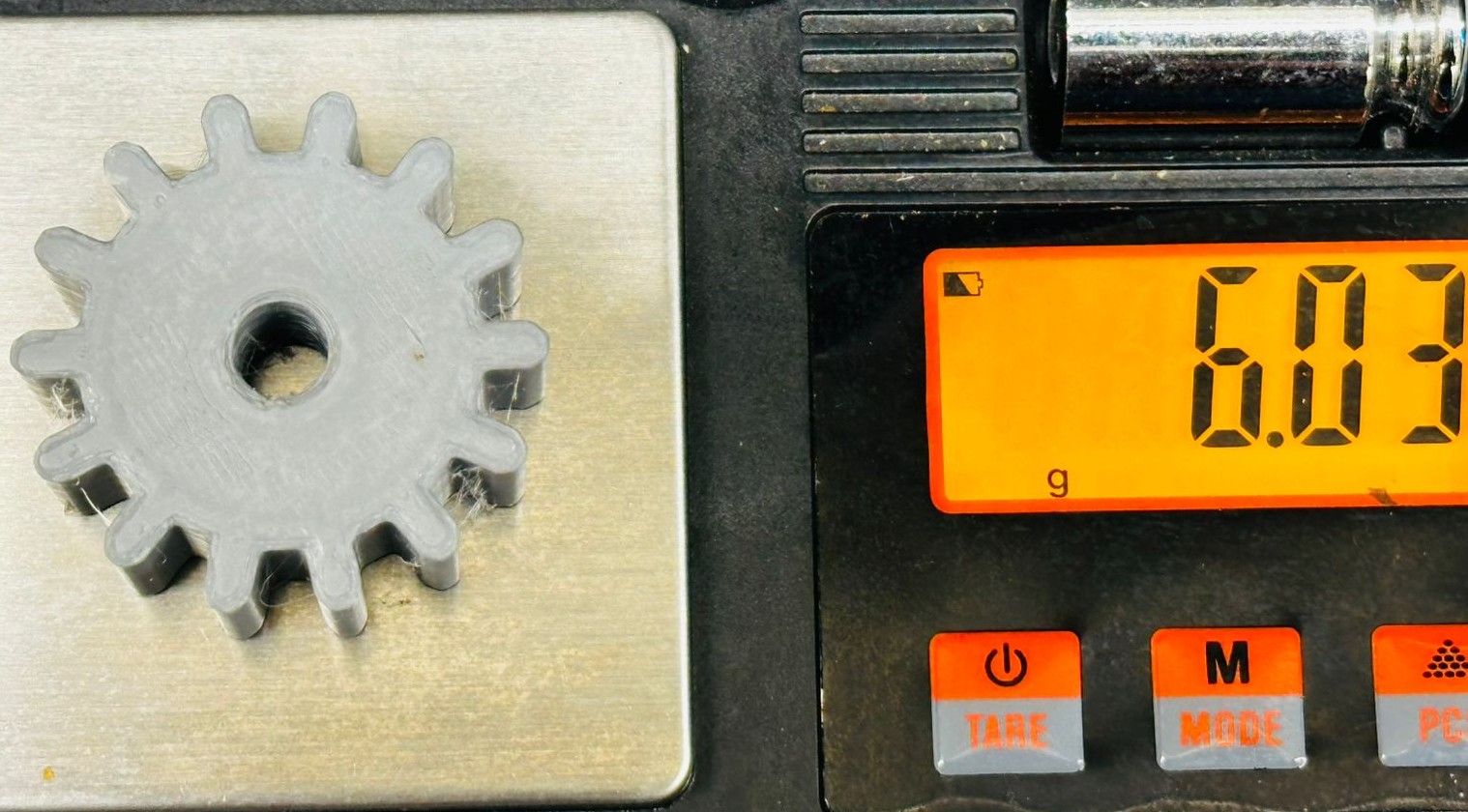}}} \hspace{0.1mm}
    \subfloat[Under-Extrusion]{\fbox{\includegraphics[width=0.28\textwidth]{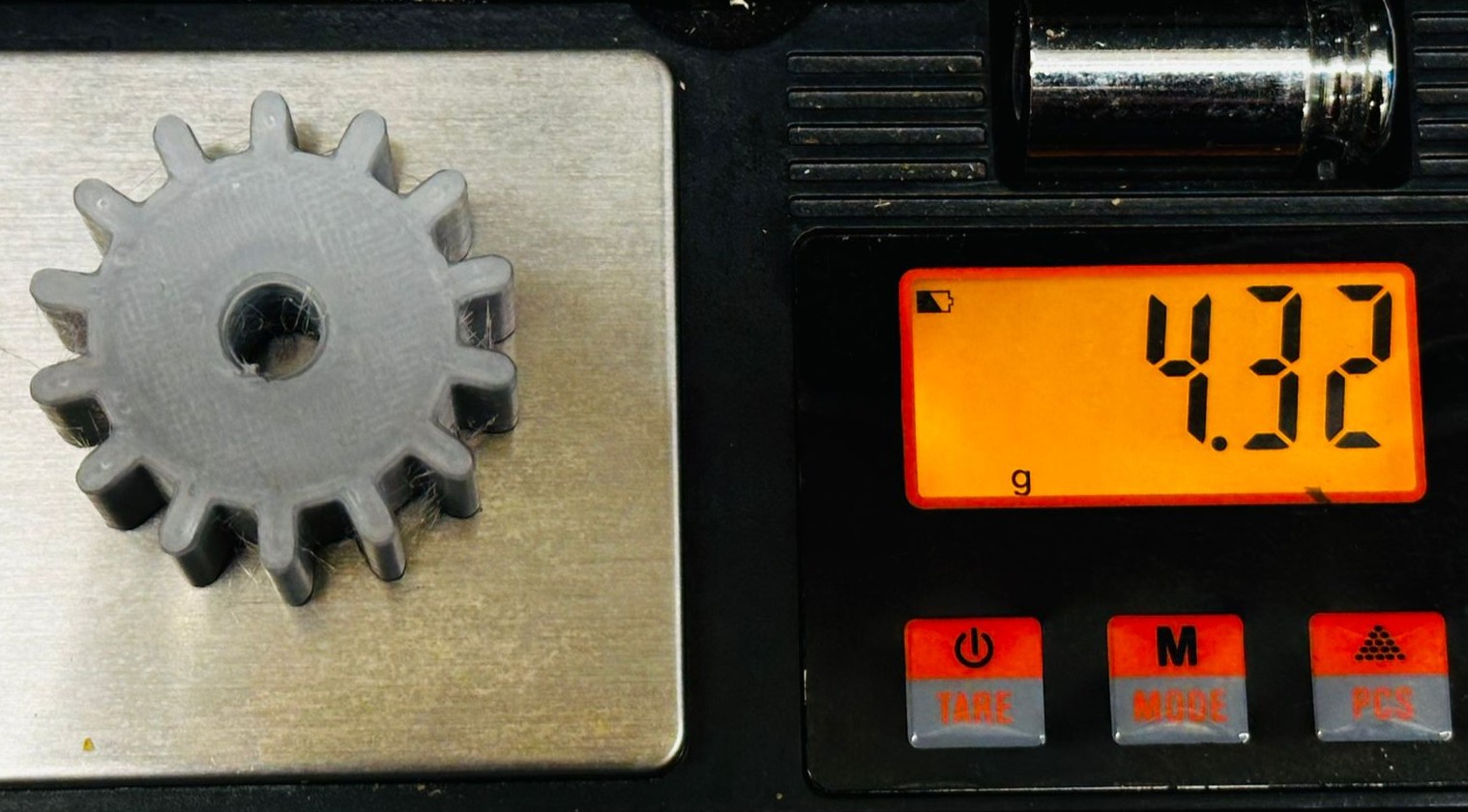}}} \hspace{0.1mm}
    \subfloat[Over-Extrusion]{\fbox{\includegraphics[width=0.28\textwidth]{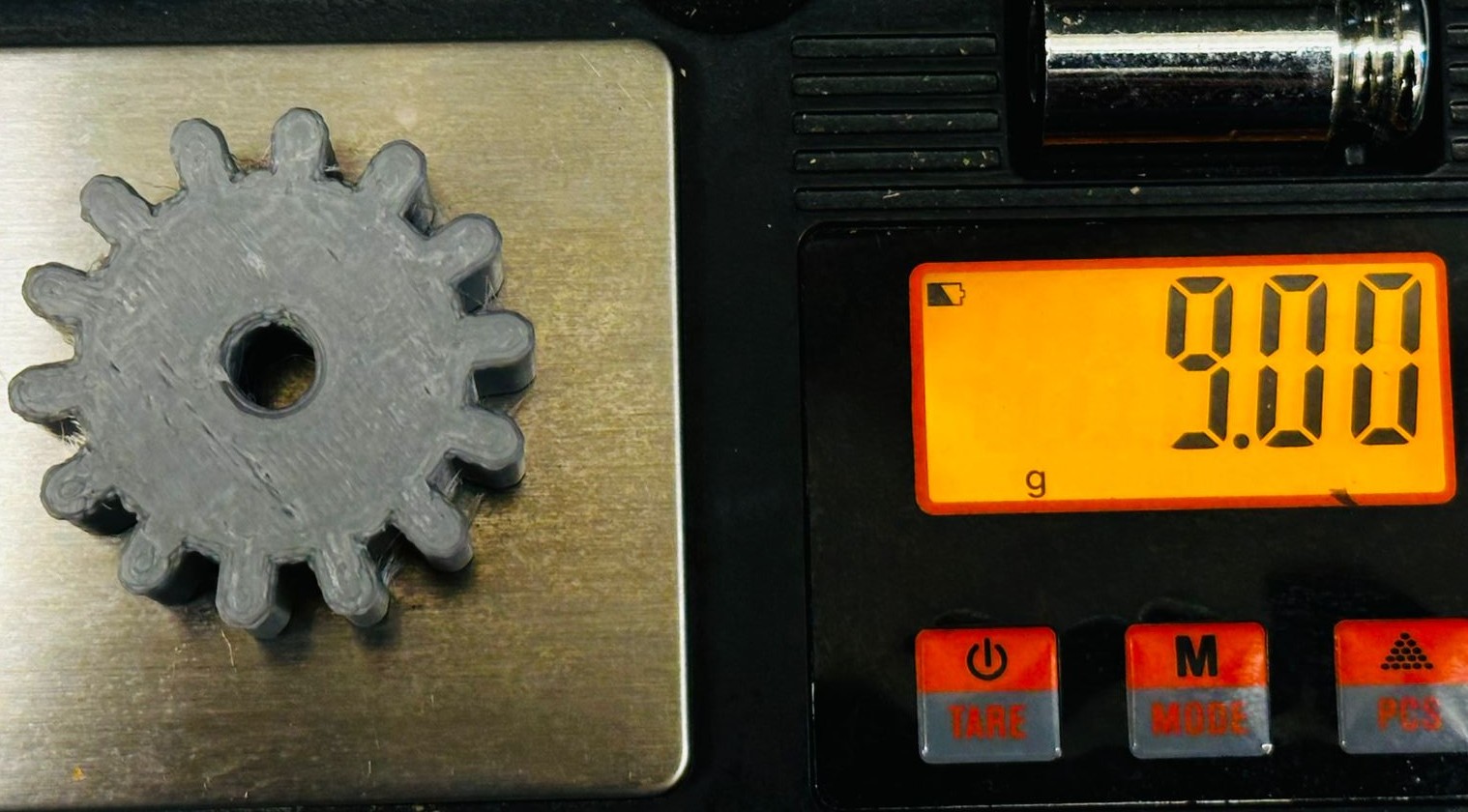}}}
    
    \caption{Creality K1-Max printed objects under different attacks.
    }
    \label{fig:K1_qual}
\end{figure}

\setlength{\fboxsep}{.5pt}
\begin{figure}[h]
    \centering
    \subfloat[Benign.]{\fbox{\includegraphics[width=0.22\textwidth]{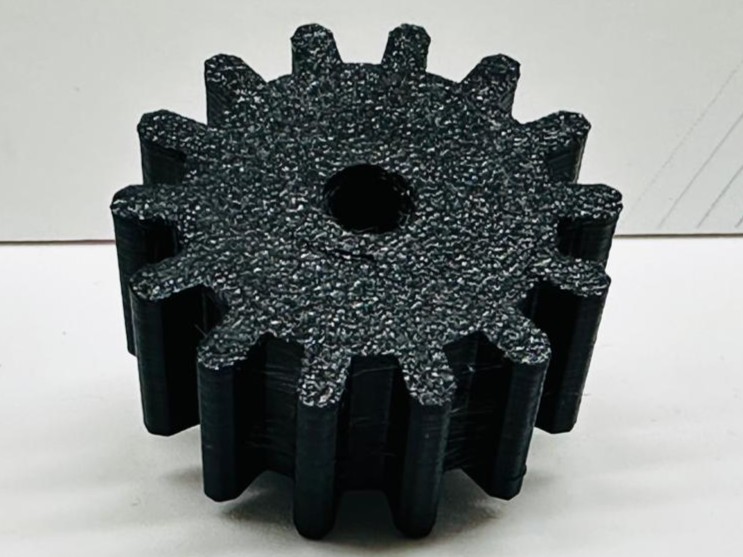}}} 
    \subfloat[Cavity Ins.]{\fbox{\includegraphics[width=0.22\textwidth]{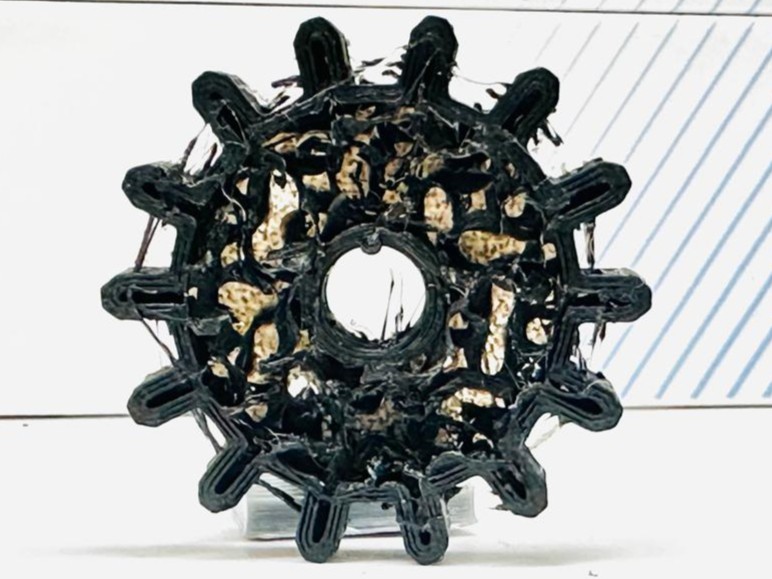}}} 
    \subfloat[Noise Inj.]{\fbox{\includegraphics[width=0.22\textwidth]{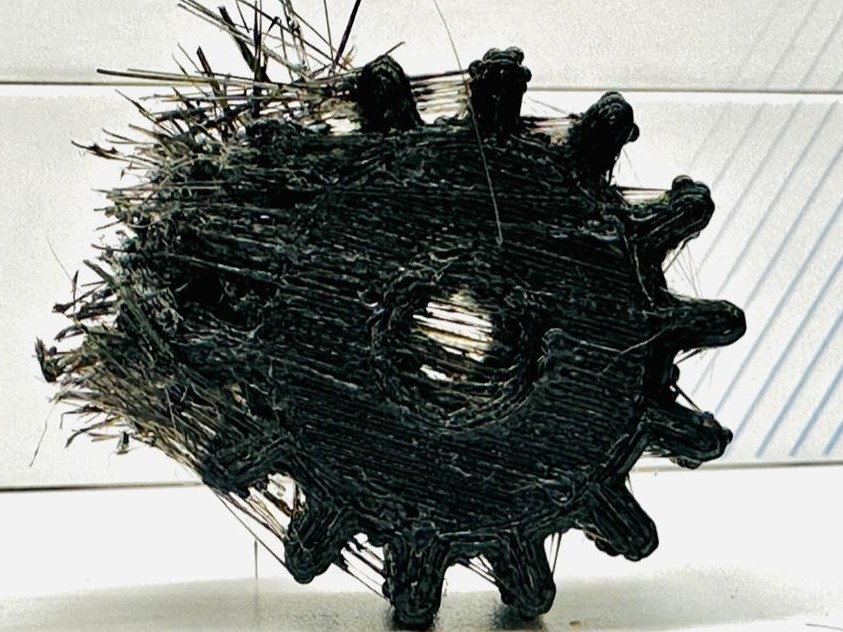}}} 
    \subfloat[Dimension Inj.]{\fbox{\includegraphics[width=0.22\textwidth]{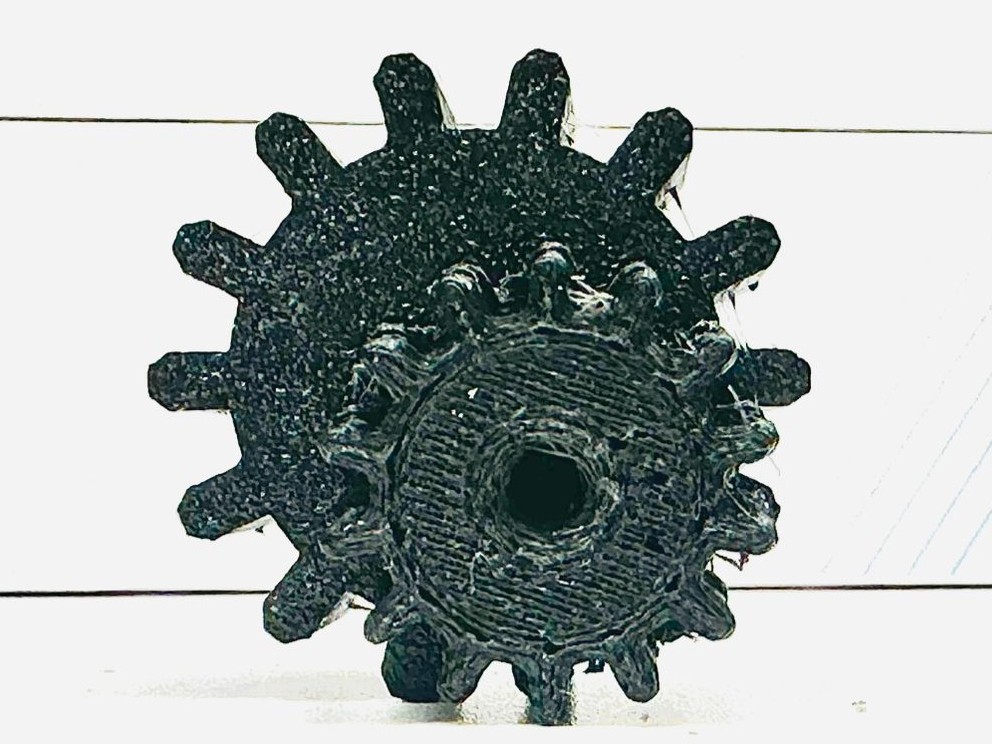}}}\\[.01mm]
    
    \subfloat[Benign Extrusion]{\fbox{\includegraphics[width=0.28\textwidth]{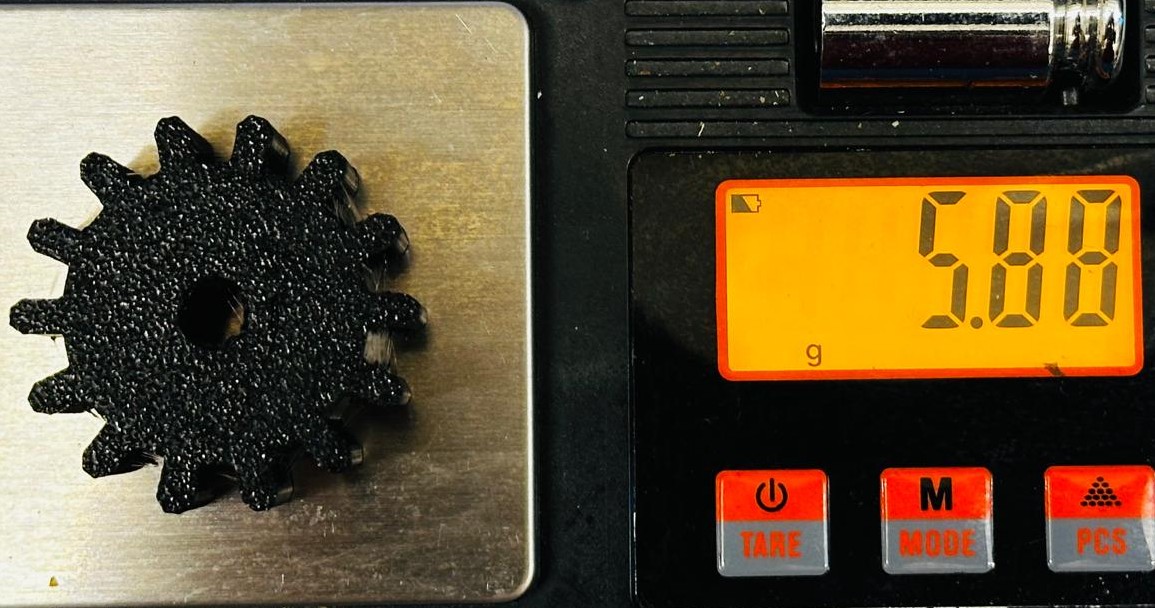}}} \hspace{0.1mm}
    \subfloat[Under-Extrusion]{\fbox{\includegraphics[width=0.28\textwidth]{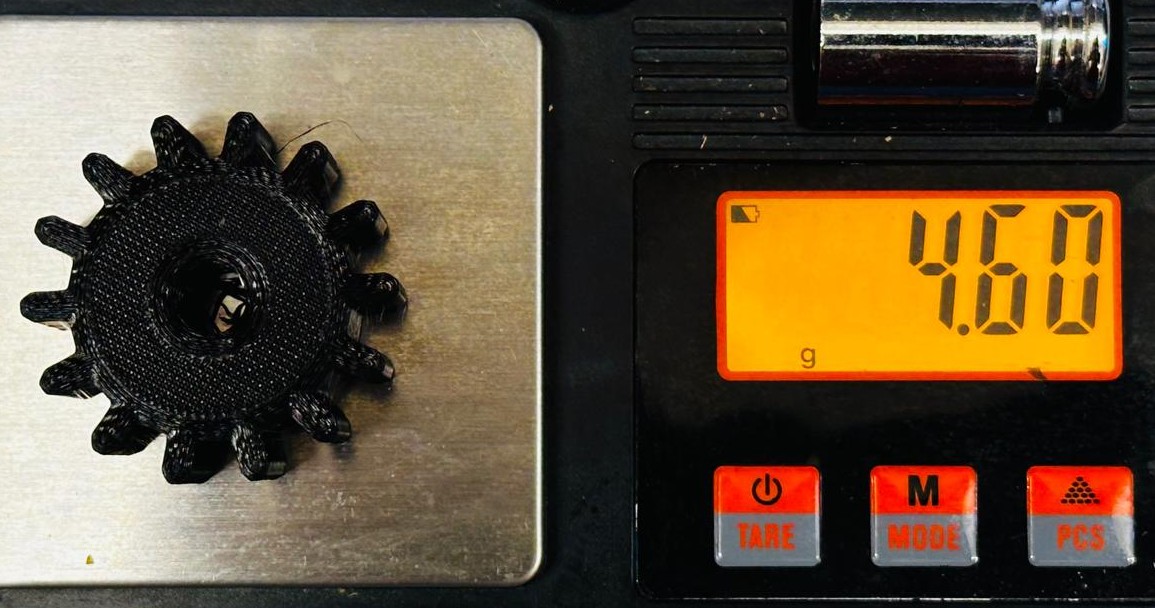}}} \hspace{0.1mm}
    \subfloat[Over-Extrusion]{\fbox{\includegraphics[width=0.28\textwidth]{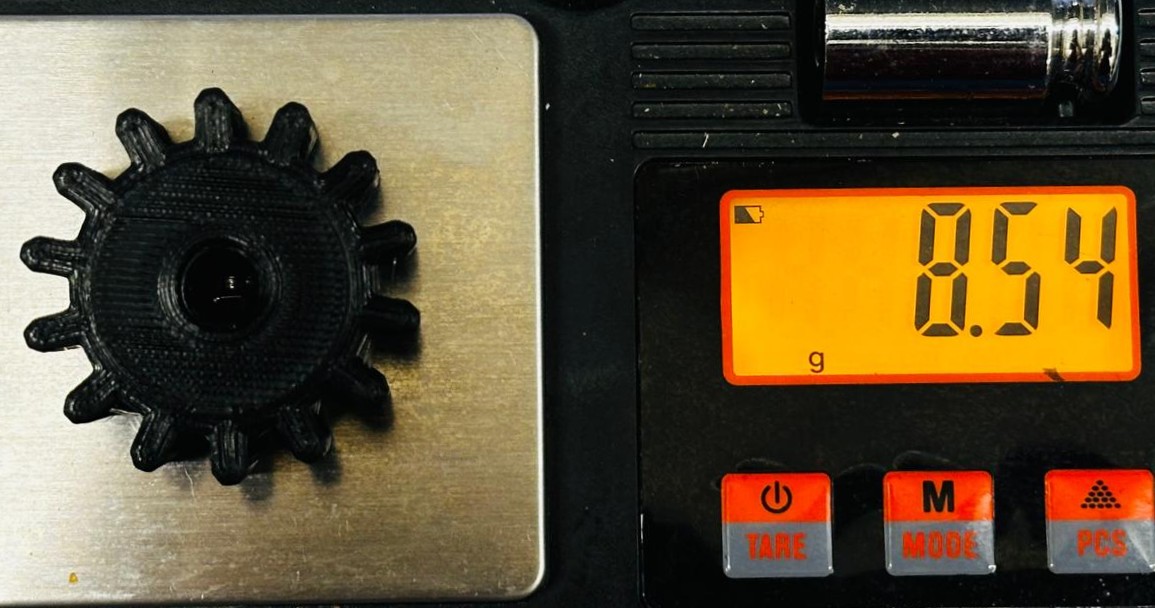}}}
    
    \caption{Ender-3 printed objects under different attacks.
    }
    \label{fig:ender_qual}
\end{figure}

Figures~\ref{fig:K1_qual} and~\ref{fig:ender_qual} illustrate the visual impact of various attack scenarios on physical objects printed using Creality K1 Max and Ender-3 printers, respectively. Each figure presents a side-by-side comparison of benign and manipulated outputs, highlighting both geometric deformation and weight anomalies.In both setups, the benign injection (Fig.\ref{fig:K1_qual}a,\ref{fig:ender_qual}a) demonstrates smooth surface finish, uniform gear teeth, and structurally complete prints serving as a visual baseline for integrity. Under the cavity insertion attack (b), both printers produce prints with missing internal material and incomplete infill, severely compromising mechanical strength. Noise injection (c) leads to stringing, irregular perimeter paths, and chaotic surface textures, revealing disrupted extrusion paths caused by G-code perturbation. Dimensional manipulation (d) visually mimics benign geometry. Still, it subtly alters the overall scale (if we scale 98\% of the original, it can't be detected by the naked eye), potentially impairing fit and function in mechanical assemblies. The lower rows focus on extrusion-level anomalies captured via precise weight measurements. For Creality K1 Max, the benign extrusion sample (e) weighs 6.03g, while under-extrusion (f) drops to 4.32g, indicating sparse filament deposition and weak part strength. Over-extrusion (g) increases the mass to 9.00g, resulting in excessive material and surface blobbing. Ender-3 results follow similar trends, benign extrusion (e) at 5.80g, under-extrusion (f) at 4.60g, and over-extrusion (g) at 8.54g, affirming that extrusion attacks manifest consistently across different printer architectures.

\section{Quantitative Evaluation}

\subsection{Clustering Results}

\begin{figure}[h]
    \centering
    \begin{subfigure}[a]{0.48\textwidth}
        \centering
        \includegraphics[width=\textwidth]{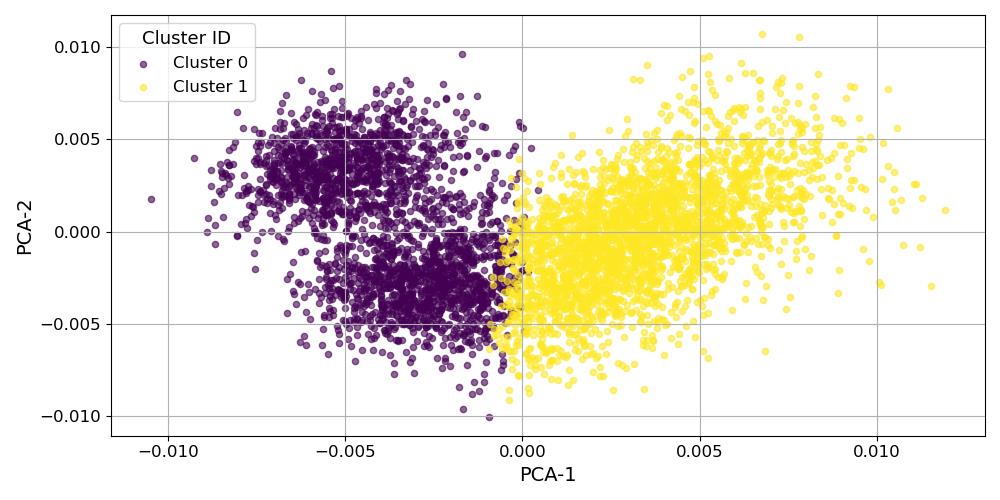}
        \caption{K-Means on PCA}
        \label{fig:pca_cluster}
    \end{subfigure}
    \hfill 
    \begin{subfigure}[a]{0.48\textwidth}
        \centering
        \includegraphics[width=\textwidth]{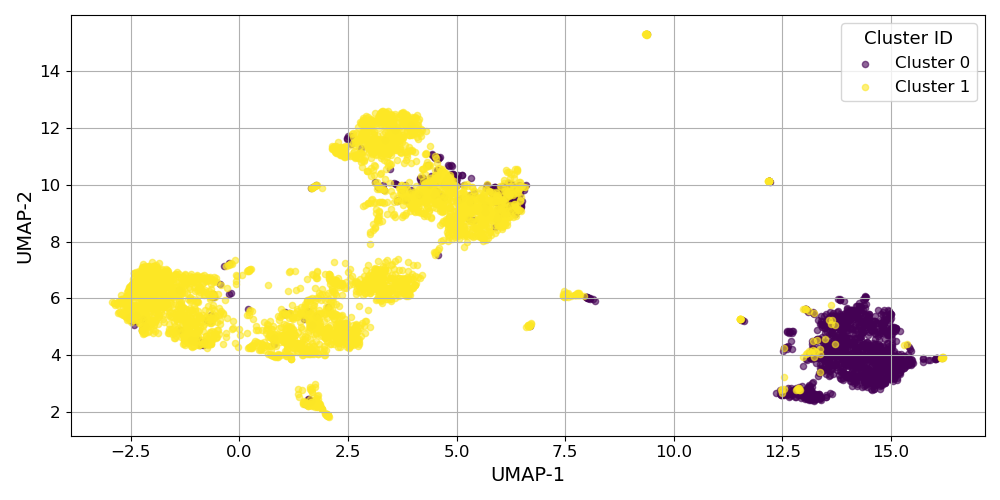}
        \caption{K-Means clustering on UMAP.}
        \label{fig:umap2d}
    \end{subfigure}
    \hfill 
    \caption{Cluster Visualization on Latent Space}
    \label{fig:PCA&UMAP}
\end{figure}

Figure~\ref{fig:pca_cluster} depicts the K-Means clustering output overlaid on a 2D PCA projection of the learned embeddings. The two clusters, shown in yellow and purple, exhibit a strong linear separation along the PCA-1 axis. This clear partitioning demonstrates that the contrastive learning framework has successfully embedded benign and anomalous behaviors into distinct regions of the latent space. 

To capture potential nonlinear structure in the learned embeddings, we employed UMAP to reduce the latent space to two dimensions. As shown in Figure~\ref{fig:umap2d}, the K-Means clusters are projected into a nonlinear manifold where the separation is even more pronounced compared to PCA. One cluster forms a tightly concentrated group on the right, while the other spans multiple regions, indicating the model’s ability to differentiate benign behaviors from diverse attack patterns. 

\subsection{Classification- Reconstruction Error:}

To differentiate benign from malicious inputs, a thresholding mechanism was employed. Specifically, we used the 95th percentile of reconstruction errors computed on benign validation data to set a threshold of 0.0019. This ensures that only the most anomalous deviations are flagged while minimizing false positives, a critical requirement in practical deployment.

\begin{table}[ht]
\small
\centering
\caption{Anomaly Detection Performance}
\label{tab:detection_metrics}
\begin{tabular}{|l|c|c|c|c|}
\hline
\multirow{2}{*}{\textbf{Metric}} & \multicolumn{2}{c|}{\textbf{Benign}} & \multicolumn{2}{c|}{\textbf{Attack}} \\
\cline{2-5}
& Value & (\%) & Value & (\%) \\
\hline
Precision       & 0.9801 & 98.01\% & 0.9055 & 90.55\% \\
Recall          & 0.9499 & 94.99\% & 0.9614 & 96.14\% \\
F1-score        & 0.9648 & 96.48\% & 0.9326 & 93.26\% \\
\hline
\multicolumn{2}{|l|}{\textbf{Overall Accuracy}}          & \multicolumn{3}{c|}{0.9537 (95.37\%)} \\
\multicolumn{2}{|l|}{\textbf{Macro Avg F1-score}}        & \multicolumn{3}{c|}{0.9487 (94.87\%)} \\
\multicolumn{2}{|l|}{\textbf{Weighted Avg F1-score}}     & \multicolumn{3}{c|}{0.9541 (95.41\%)} \\
\multicolumn{2}{|l|}{\textbf{AUROC (AUC)}}               & \multicolumn{3}{c|}{0.9870 (98.70\%)} \\
\multicolumn{2}{|l|}{\textbf{Threshold (MSE)}}           & \multicolumn{3}{c|}{0.0019} \\
\hline
\end{tabular}
\end{table}

The model's performance was quantitatively evaluated using a confusion matrix, ROC curve, and classification metrics in a Table~\ref{tab:detection_metrics}. The self-autoencoder achieved a high overall classification accuracy of 95.37\%, with a precision of 90.55\%, recall of 96.14\%, and an F1-score of 93.26\%. Notably, the area under the ROC curve (AUC) 
was 0.98, demonstrating strong discriminative power between benign and anomalous inputs.

\begin{table}[h]
\small
\centering
\caption{Autoencoder Output}
\label{tab:confmat_ae}
\begin{tabular}{|c|c|c|}
\hline
\textbf{True / Predicted} & \textbf{Benign} & \textbf{Attack} \\
\hline
\textbf{Benign} & 94.99\% & 5.01\% \\
\textbf{Attack} & 3.86\% & 96.14\% \\
\hline
\end{tabular}
\end{table}

The confusion matrix in Table~\ref{tab:confmat_ae} further reveals that 94.99\% of benign samples were correctly classified as benign, while only 5.01\% were misclassified as attacks. For attack samples, 96.14\% were correctly identified as malicious, whereas 3.86\% were incorrectly marked as benign. 


\begin{figure}[h]
    \centering
    \begin{subfigure}[a]{0.5\textwidth}
        \centering
        \includegraphics[width=\textwidth]{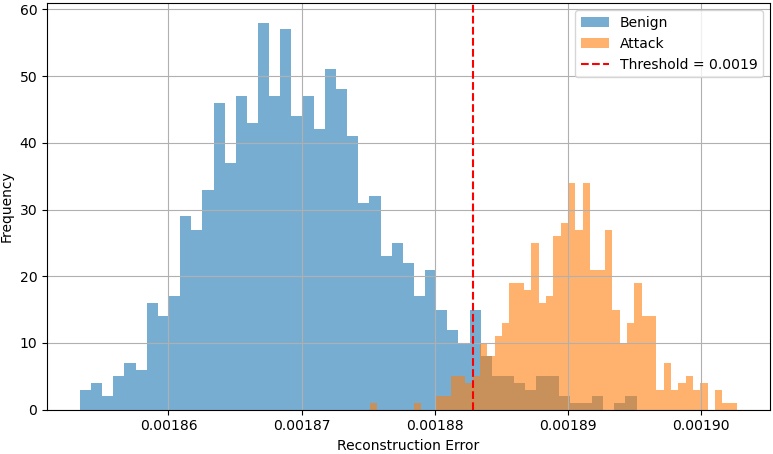}
        \caption{Histogram of Reconstruction Error}
        \label{fig:reconstruction_hist}
    \end{subfigure}
    \hfill 
    \begin{subfigure}[a]{0.6\textwidth}
        \centering
        \includegraphics[width=\textwidth]{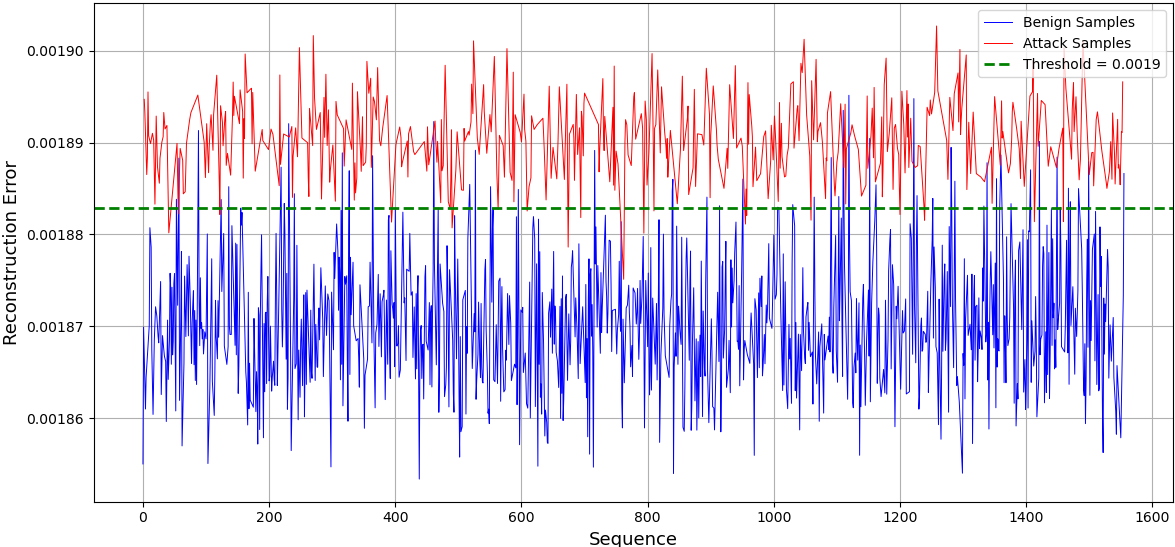}
        \caption{Reconstruction error by Sequence.}
        \label{fig:reconstruction_sequence}
    \end{subfigure}
    \hfill 
    \caption{Visualization of Reconstruction Errors(MSE)}
    \label{fig:thrsplot}
\end{figure}

We further visualized the reconstruction errors for benign and attack samples. Figure~\ref{fig:reconstruction_hist} presents the distribution of reconstruction errors for each class. Benign samples (in blue) exhibit a tightly grouped error distribution centered around a lower mean, indicating high fidelity in reconstruction. In contrast, attack samples (in orange) display a right-shifted distribution with a larger spread. 


Figure~\ref{fig:reconstruction_sequence} further corroborates these findings by illustrating reconstruction error across the sequence of test samples. Benign samples (blue line) consistently maintain lower reconstruction errors below the defined threshold. Conversely, attack samples (red line) frequently cross this boundary, suggesting pronounced deviations from the learned benign patterns. 

\section{Conclusion}
In this study, we demonstrate how various attack vectors, including G-code hijacking, cavity insertion, under- or over-extrusion, dimensional manipulation, and intellectual property theft, can be initiated through backdoor access or mid-print tampering. We also developed and evaluated an unsupervised intrusion detection framework that uses contrastive representation learning and structured log analysis. Our methodology learns semantic embeddings of benign printer behavior and employs both clustering-based and reconstruction-based strategies to detect deviations. Our experiments on Creality K1 Max and Ender 3 printers validated the feasibility of these attacks and the efficacy of our defense model.

Despite its effectiveness, the current system has some limitations. First, the approach assumes access to structured and timestamp-aligned logs, which may not be available in proprietary or closed-source firmware environments. Second, specific stealthy attacks that do not significantly perturb telemetry patterns (e.g., subtle IP theft without extrusion anomalies) may remain undetected. 
Future directions include incorporating temporal modeling using graph-based temporal embeddings to capture long-range dependencies. 
Finally, generalizing the model across multiple printer architectures and expanding to federated anomaly detection across a distributed fleet of printers will broaden applicability.

\section{Acknowledgment}
This material is based upon work supported by the National Science Foundation under Award No. 2417062. Any opinions, findings, and conclusions or recommendations expressed in this material are those of the authors and do not necessarily reflect the views of the National Science Foundation.

\bibliographystyle{unsrt}
\bibliography{ref}

\end{document}